\documentstyle[12pt,epsfig]{article}
\def\lsim{\raise0.3ex\hbox{$<$\kern-0.75em\raise-1.1ex\hbox{$\sim$}}}
\def\gsim{\raise0.3ex\hbox{$>$\kern-0.75em\raise-1.1ex\hbox{$\sim$}}}

\def\beq{\begin{equation}}
\def\eeq{\end{equation}}
\def\bea{\begin{eqnarray}}
\def\eea{\end{eqnarray}}
\def\bq{\begin{quote}}
\def\eq{\end{quote}}

\newcommand{\rr}{\mbox{\boldmath $r$}}

\def\gappeq{\mathrel{\rlap {\raise.5ex\hbox{$>$}}
{\lower.5ex\hbox{$\sim$}}}}

\def\lappeq{\mathrel{\rlap{\raise.5ex\hbox{$<$}}
{\lower.5ex\hbox{$\sim$}}}}

\def\Toprel#1\over#2{\mathrel{\mathop{#2}\limits^{#1}}}

\newcommand{\rk}{\mbox{\boldmath $k$}}

\begin{document}
\pagestyle{empty}
\begin{center}
  {\bf THE QCD POMERON IN ULTRAPERIPHERAL HEAVY ION COLLISIONS: IV.  PHOTONUCLEAR PRODUCTION OF VECTOR MESONS}
\\

\vspace*{1cm}
 V.P. Gon\c{c}alves $^{1}$, M.V.T. Machado  $^{2,\,3}$\\
\vspace{0.3cm}
{$^{1}$ Instituto de F\'{\i}sica e Matem\'atica,  Universidade
Federal de Pelotas\\
Caixa Postal 354, CEP 96010-090, Pelotas, RS, Brazil\\
$^{2}$ \rm Universidade Estadual do Rio Grande do Sul - UERGS\\
 Unidade de Bento Gon\c{c}alves. CEP 95700-000. Bento Gon\c{c}alves, RS, Brazil\\
$^{3}$ \rm High Energy Physics Phenomenology Group, GFPAE  IF-UFRGS \\
Caixa Postal 15051, CEP 91501-970, Porto Alegre, RS, Brazil}\\
\vspace*{1cm}
{\bf ABSTRACT}
\end{center}

\vspace*{1.7cm} \noindent

\vspace*{1.5cm} \noindent \rule[.1in]{17cm}{.002in}

\vspace{-5cm} \setcounter{page}{1} \pagestyle{plain} 

The photonuclear production of vector mesons in ultraperipheral heavy ion collisions is investigated within the QCD color dipole picture, with 
 particular emphasis on the saturation model. The integrated cross section and the rapidity distribution for the $AA \rightarrow V\,AA$ ($V = \rho, \omega, \phi, J/\Psi$) process are computed and theoretical estimates for scattering on both  light and heavy nuclei are given for energies of RHIC and LHC. A comparison with the recent STAR data on coherent production of $\rho$ mesons is also presented. Furthermore, we calculate the photoproduction of vector mesons in proton-proton collisions at RHIC, Tevatron and LHC energies.  \vspace{0.5cm}

\section{Introduction}

In ultraperipheral relativistic heavy-ion collisions (UPC's) the ions do
not interact directly with each other and move essentially
undisturbed along the beam direction. The only possible
interaction is due to the long range electromagnetic interaction
and diffractive processes (For a review see, e. g. Ref.
\cite{bert}). Due to the coherent action of all the protons in the
nucleus, the electromagnetic field is very strong and the
resulting flux of equivalent photons is large. A photon stemming
from the electromagnetic field of one of the two colliding nuclei
can penetrate into the other nucleus and interact with one or more
of its hadrons, giving rise to photon-nucleus collisions to an
energy region hitherto unexplored experimentally. For example, the
interaction of quasi-real photons with protons has been studied
extensively at the electron-proton collider at HERA, with
$\sqrt{s} = 300$ GeV. The obtained $\gamma p$ center of mass
energies extends up to $W_{\gamma p} \approx 200$ GeV, an order of
magnitude larger than those reached by fixed target experiments.
Due to the larger number of photons coming from one of the
colliding nuclei in heavy ion collisions, a  similar and more detailed
study will be possible in these collisions, with $W_{\gamma N}$
reaching 950 GeV for the Large Hadron Collider (LHC) operating in
its heavy ion mode. Similarly, the strong electromagnetic fields generated by high-energy protons allow us to study photon - nucleon processes in proton - proton interactions in a large kinematical range. These events can be experimentally studied by selecting events with low multiplicity and small total transverse momentum.

Over the past few  years a comprehensive analysis of the heavy quark \cite{antigos,vicber,klein_vogt,per2,per3} and vector meson \cite{vicber,klein_nis_prc,strikman,per1,double_meson,klein_nis_prl} production in ultraperipheral heavy ion collisions was made considering different theoretical approaches. In particular,   much effort has been devoted to obtain signatures of the QCD Pomeron in such processes \cite{per2,per3,per1}, which can be used to constrain the QCD dynamics at high energies.
Understanding the behavior of high energy hadron reactions from a
fundamental perspective within QCD is an important goal of
particle physics. In the late 1970s, Lipatov and
collaborators \cite{bfkl} established the papers which form the
core of our knowledge of Regge limit (high energy limit) of
QCD. The physical effect that they
describe is often referred to as the QCD Pomeron, with the
evolution described by the BFKL equation. Since this equation predicts
that for $s \rightarrow \infty$ the corresponding cross section
rises with a  power law of the energy, violating the Froissart
bound,  new dynamical effects associated with the unitarity
corrections are expected to stop further growth of the cross
sections \cite{glr}. This expectation can be easily understood:
while for large transverse momentum $k_{\perp}$, the BFKL equation
predicts that the mechanism $g \rightarrow gg $ populates the
transverse space with a large number of small size gluons per unit
of rapidity (the transverse size a gluon with momentum $k_{\perp}$ is
proportional to $1/k_{\perp}^2$), for small $k_{\perp}$ the produced gluons
overlap and fusion processes, $gg \rightarrow g$, are equally
important. Considering this process,  the rise of the gluon
distribution with transverse momenta below a typical scale, energy
dependent, called saturation scale $Q_{\mathrm{sat}}$ is reduced, restoring the
unitarity. It is important to emphasize that Golec-Biernat and
Wusthoff \cite{golecwus} have shown that a saturation model  is able
to describe the DESY $ep$ collider HERA data, in particular
the transition from the perturbative  to the nonperturbative
photoproduction region (For improvements of this model see Refs. \cite{bgbk,kowtea}). Moreover, its parameter-free application to diffractive DIS has been
also quite successful \cite{golecwus} as well as its extension to
virtual Compton scattering \cite{Favart_Machado}, vector meson
production \cite{mara,stasto,sandapen}, charm and longitudinal structure function \cite{vicmag_prl, vicmag_fl}  and two-photon collisions
\cite{Kwien_Motyka}.   However, once other approaches containing very distinct assumptions also describe the same set of HERA data,  
the description of the QCD Pomeron
still is an open question and it deserves more detailed analyzes.  An alternative way to constrain the QCD dynamics are studies on  electron - nucleus interactions, since the saturation effects are amplified  in a nuclear medium - $Q_{s\,A}^2 \propto A^{1/3}$. Several works have been made along these lines, studying the behavior of the observables in the kinematic region of the planned $eA$ colliders at HERA and RHIC (For recent reviews see, e.g., Refs. \cite{iancu_venu,vic_jbras}). In particular,  
 recently we have analyzed the nuclear exclusive vector meson production  considering two distinct theoretical scenarios \cite{magno_victor_mesons}.   
Our results have demonstrated that the experimental analyzes of  nuclear
exclusive vector meson photoproduction in the future
electron-nucleus colliders eRHIC and HERA-A could be useful to
discriminate between the different theoretical scenarios, mainly
if heavy nuclei are considered.
It strongly motivates the extension of these studies for ultraperipheral heavy ion collisions which can be analyzed in the current and/or scheduled colliders.

Recently, the STAR Collaboration released the first data on the cross section of the coherent $\rho$ production in gold - gold ultraperipheral collisions at $\sqrt{s} = 130$ GeV \cite{star_data}, providing the first opportunity to check the basic features and main approximations of the distinct approaches describing nuclear vector meson photoproduction.
The main aspect is that  real photons have a complicated nature. In a first approximation, the photon is a point-like particle, although in field theory it may fluctuate also into a fermion pair (See discussions in Refs. \cite{bert,nisius}). In the case where  there is a  photon transition in a colorless antiquark-quark pair, the propagation of this colorless hadronic wave packet in a nuclear medium can be treated either in the hadronic basis as a result of Gribov's inelastic corrections or in QCD in terms of the partonic basis, which are complementary. Lets briefly discuss  these two representations (For a detailed discussion see Ref. \cite{Nikolaev}).  The time scale characterizing the evolution of a $q \overline{q}$ wave packet can be estimated based on the uncertainty principle and Lorentz dilation. The lifetime of 
the photon fluctuation is given by $t_c =\nu/(Q^2 +  m^2_V)$,
where  $\nu$ is the photon energy, $m_V$ is the mass of the fluctuation and $Q^2$ is the photon virtuality. It is usually called coherence time. Using light-cone kinematics we can define the coherence length, which is given by $l_c = t_c$. Moreover, one cannot decide whether a ground state $V$ is produced or the next-excited state $V^{\prime}$, unless the process lasts longer than the inverse mass difference between these states. In the rest frame of the nucleus, this formation time is Lorentz dilated and is given by $t_f = 2 \nu/(m^2_{V^{\prime}} - m^2_V)$.  Similarly, we can define a formation length given by $l_f = t_f$. In the hadronic basis, the same process looks quite different. The incident photon may produce different states on a bound nucleon,  the $V$ meson ground state or an excited state. Those states propagate through the nucleus experiencing multiple-diagonal and off-diagonal diffractive interactions, and eventually the ground state is detected. According the quark-hadron duality, we expect that these two descriptions   to be equivalent. However, as these two approaches have been used assuming different approximations, their comparison may provide a scale for the theoretical uncertainty involved.   
Furthermore, it is important to emphasize that at high photon energy $\nu$, both $l_c$ and $l_f$ greatly exceed the nuclear radius $R_A$, which implies in the partonic basis that the transverse size of the $q\overline{q}$ pair do not change during the interaction with the target. This enables one to introduce the QCD dipole picture \cite{nik}, where the process is factorized into the photon fluctuation in a $q\overline{q}$ pair and the dipole cross section.  These aspects become the interpretation in the partonic basis more intuitive and straightforward than in the hadronic basis. Since the photoproduction of vector mesons in ultraperipheral heavy ion collisions have been analyzed in literature using only the hadronic basis, it motivates the study of this process using the partonic one. 

Other important point which motivates our analysis is that in the current studies of  vector meson production in UPC's the extrapolations of the predictions for LHC energies in general assume a power-like behavior for the $\gamma p \,(A) \rightarrow V p \,(A)$ cross sections. Despite the good agreement for the currently  available energies, an extrapolation to higher energies of the 
experimental fits implies a large growth for the cross section
which it would violate the unitarity at sufficiently high energies. Therefore,
dynamical modifications associated to the unitarity corrections
are also expected to be present in vector meson production \cite{mara,stasto,sandapen}. As discussed before, these effects should be enhanced in nuclear
processes \cite{glr,levin,magno_victor_mesons}.

In this paper the photonuclear production of vector mesons in UPC's is investigated within the QCD color dipole picture \cite{nik}, with  particular emphasis on the saturation model \cite{golecwus}. The exclusive vector meson production by real and virtual photons is
 an outstanding process  providing important information on the transition
region from the soft dynamics (at low virtualities of the photon
$Q^2$) to the hard perturbative regime at high $Q^2$
\cite{predazzi,crittenden} (For a recent review see Ref. \cite{nik_iva}). In principle, a perturbative approach
is only justified if a hard scale is present in the process, e.g. the
photon virtuality and/or a large mass of the vector meson. For photoproduction of light mesons,
 such scale is not present and one has to rely on non-perturbative models.  On some pQCD approaches, as the saturation model, even this soft process can be described, where the transition is set by the saturation scale. 
 In the color dipole approach, the degrees of freedom are
the photon (color dipole) and meson wavefunctions as well as the
dipole-nuclei cross section. Such an approach enables to include
nuclear effects and parton saturation phenomenon. The latter one is
characterized by the typical momentum (saturation) scale $Q_{\mathrm{sat}}$, which it has been constrained by
experimental results in deep inelastic scattering (DIS) and
diffractive DIS \cite{golecwus}. Here, we will use an extension of the phenomenological saturation model for  nuclear
targets \cite{armesto}. This model  reasonably describes  the
experimental data for the nuclear structure function  and has been
used to predict the nuclear inclusive and diffractive cross
sections for heavy quark photoproduction \cite{vicmagsat}. The nuclear saturation scale,
$Q_{\mathrm{s}\,A}$, provides the transition between the
color transparency  and the saturation regimes in the nuclear scattering. Concerning vector meson production, our starting point are
the recent works in Refs. \cite{sandapen,magno_victor_mesons}, where this approach was applied for the
proton and nuclear case. It is worth mentioning that although light meson
photoproduction to be a soft process by definition, it is
consistently described in the QCD color dipole picture whether
there is a suitable model for the soft-hard transition, as occurring in the 
the saturation model.
As a by product, we extend our analysis for the photoproduction of vector mesons in proton-proton collisions at RHIC, Tevatron and LHC energies.

This paper is organized as  follows. In the next section, we present a brief review of the
 ultraperipheral heavy ion collisions and the main formulae to describe the photon - hadron process in these  reactions. Moreover, the photon spectrum 
in proton - proton collisions is discussed.   
In Section \ref{models} we discuss the photoproduction of vector mesons in the QCD color dipole picture and the saturation model is shortly reviewed.   In Section
\ref{discussion} we present our results for the integrated cross section and the rapidity distribution for the $AA \rightarrow AAV$  and  $pp \rightarrow ppV$processes, where $V = \rho, \omega, \phi, J/\Psi$. Moreover, it is presented a  comparison of our predictions with the STAR data for coherent $\rho$ photoproduction on nuclei at energy $\sqrt{s_{NN}}=130$ GeV and a discussion concerning related approaches used in its description is addressed.  Finally, in the last section  we present a summary of the main results and conclusions.

\section{Photonuclear vector meson production at UPC's}

In heavy ion  collisions the large number of photons coming from
one of the colliding nuclei  will  allow to study photoproduction,
with energies $W_{\gamma N}$ reaching to  950 GeV for the LHC. The
photonuclear cross sections are given by the convolution between
the photon flux from one of the nuclei and the cross section for
the scattering photon-nuclei. 
The final  expression
for the production of vector mesons in ultraperipheral heavy ion
collisions is then given by,
\begin{eqnarray}
\sigma_{AA \rightarrow AAV}\,\left(\sqrt{S_{\mathrm{NN}}}\right) = \int \limits_{\omega_{min}}^{\infty} d\omega \, \frac{dN\,(\omega)}{d\omega}\,\, \sigma_{\gamma \,A \rightarrow VA} \left(W_{\gamma A}^2=2\,\omega\sqrt{S_{\mathrm{NN}}}\right)\,
\label{sigAA}
\end{eqnarray}
where $\omega$ is the photon energy  with $\omega_{min}=m_V^2/4\gamma_L m_p$ and
$\sqrt{S_{\mathrm{NN}}}$ is  the ion-ion c.m.s energy. The Lorentz factor for LHC is
$\gamma_L=2930$, giving the maximum c.m.s. $\gamma N$ energy
$W_{\gamma A} \lappeq 950$ GeV.  In this process we have that the nuclei are not disrupted and the final state consists solely of the two nuclei and the vector meson decay products. Consequently, we have that the final state is  characterized by a small number of centrally produced particles, with rapidity gaps separating the central final state from both beams. Moreover, due to the coherence requirement, the transverse momentum  is limited to be smaller than $p_T = \sqrt{2}/R_A$, where $R_A$ is the nuclear radius. Therefore, these reactions can be studied experimentally by selecting events with low multiplicity and small total $p_T$.

The photon flux is given by the
Weizsacker-Williams method \cite{bert}. The flux from a charge
$Z$ nucleus a distance $b$ away is
\begin{eqnarray}
\frac{d^3N\,(\omega,\,b^2)}{d\omega\,d^2b}= \frac{Z^2\alpha_{em}\eta^2}{\pi^2 \,\omega\, b^2}\, \left[K_1^2\,(\eta) + \frac{1}{\gamma_L^2}\,K_0^2\,(\eta) \right] \,
\label{fluxunint}
\end{eqnarray}
where $\gamma_L$ is the Lorentz boost  of a single beam and $\eta
= \omega b/\gamma_L$; $K_{0,\,1}(\eta)$ are the
modified Bessel functions. The requirement that  photoproduction
is not accompanied by hadronic interaction (ultraperipheral
collision) can be done by restricting the impact parameter $b$  to
be larger than twice the nuclear radius, $R_A=1.2 \,A^{1/3}$ fm.
Therefore, the total photon flux interacting with the target
nucleus is given by Eq. (\ref{fluxunint}) integrated over the
transverse area of the target for all impact parameters subject to
the constraint that the two nuclei do not interact hadronically.
An analytic approximation for $AA$ collisions can be obtained
using as integration limit $b>2\,R_A$, producing
\begin{eqnarray}
\frac{dN\,(\omega)}{d\omega}= \frac{2\,Z^2\alpha_{em}}{\pi\,\omega}\, \left[\bar{\eta}\,K_0\,(\bar{\eta})\, K_1\,(\bar{\eta})+ \frac{\bar{\eta}^2}{2}\,\left(K_1^2\,(\bar{\eta})-  K_0^2\,(\bar{\eta}) \right) \right] \,
\label{fluxint}
\end{eqnarray}
where $\bar{\eta}=2\omega\,R_A/\gamma_L$.
It is worth mentioning that the
difference between the complete numeric and the analytical
calculation presented above  for the photon flux is less than 15
\% for the most of the purposes \cite{bert}.

In a similar way, vector meson production also occur when considering energetic protons in $pp(\bar{p})$ colliders \cite{klein_nis_prl}. In this case the  photon spectrum is given by  \cite{Drees},
\begin{eqnarray}
\frac{dN(\omega)}{d\omega} =  \frac{\alpha_{\mathrm{em}}}{2 \pi\, \omega} \left[ 1 + \left(1 - 
\frac{2\,\omega}{\sqrt{S_{NN}}}\right)^2 \right] 
\left( \ln{\Omega} - \frac{11}{6} + \frac{3}{\Omega}  - \frac{3}{2 \,\Omega^2} + \frac{1}{3 \,\Omega^3} \right) \,,
\label{eq:photon_spectrum}
\end{eqnarray}
with the notation $\Omega = 1 + [\,(0.71 \,\mathrm{GeV}^2)/Q_{\mathrm{min}}^2\,]$ and $Q_{\mathrm{min}}^2= \omega^2/[\,\gamma_L^2 \,(1-2\,\omega /\sqrt{S_{NN}})\,] \approx (\omega/
\gamma_L)^2$. The expression above is derived considering the Weizs\"{a}cker-Williams method of virtual photons and using an elastic proton form factor (For more detail see Refs. \cite{klein_nis_prl,Drees}).
It is important to emphasize that the expression (\ref{eq:photon_spectrum})  is based on a heuristic approximation,
which leads to an overestimation of the cross section at high energies ( $\approx 11 \%$ at $\sqrt{s}=1.3$ TeV)  in comparison with the more rigorous derivation of the photon spectrum for elastic scattering on protons derived in Ref. \cite{kniehl}.  For a more detailed comparison among the different photon spectra see Ref. \cite{nys_fluxo}.

\section{Vector meson production in the color dipole approach}
\label{models}

Let us consider the
scattering process $\gamma p \rightarrow Vp$ in the QCD dipole approach, where $V$ stands for
both light and heavy mesons. The scattering process can be seen
in the target rest frame as a succession in time of three
factorizable subprocesses: i) the photon fluctuates in a
quark-antiquark pair (the dipole), ii) this color dipole interacts with the
target and, iii) the pair converts into vector meson final state.
Using as kinematic variables the $\gamma^* N$ c.m.s. energy
squared $s=W_{\gamma N}^2=(p+q)^2$, where $p$ and $q$ are the target and the
photon momenta, respectively, the photon virtuality squared
$Q^2=-q^2$ and the Bjorken variable $x=Q^2/(W_{\gamma N}^2+Q^2)$, the
corresponding imaginary part of the amplitude at zero momentum
transfer reads as \cite{nik},
\begin{eqnarray}
{\cal I}m \, {\cal A}\, (\gamma p \rightarrow Vp)  = \sum_{h, \bar{h}}
\int dz\, d^2\rr \,\Psi^\gamma_{h, \bar{h}}(z,\,\rr,\,Q^2)\,\sigma_{dip}^{\mathrm{target}}(\tilde{x},\rr) \, \Psi^{V*}_{h, \bar{h}}(z,\,\rr) \, ,
\label{sigmatot}
\end{eqnarray}
where $\Psi^{\gamma}_{h, \bar{h}}(z,\,\rr)$ and $\Psi^{V}_{h,
  \bar{h}}(z,\,\rr)$  are the light-cone wavefunctions  of the photon
  and vector meson, respectively. The quark and antiquark helicities are labeled by $h$ and $\bar{h}$
  and reference to the meson and photon helicities are implicitly understood. The variable $\rr$ defines the relative transverse
separation of the pair (dipole) and $z$ $(1-z)$ is the
longitudinal momentum fractions of the quark (antiquark). The basic
blocks are the photon wavefunction, $\Psi^{\gamma}$, the  meson
wavefunction, $\Psi_{T,\,L}^{V}$,  and the dipole-target  cross
section, $\sigma_{dip}^{\mathrm{target}}$.

In the dipole formalism, the light-cone
 wavefunctions $\Psi_{h,\bar{h}}(z,\,\rr)$ in the mixed
 representation $(z,\rr)$ are obtained through two dimensional Fourier
 transform of the momentum space light-cone wavefunctions, 
 $\Psi_{h,\bar{h}}(z,\,\rk)$  which can be completely determined using light cone perturbation theory (see more details, e.g. in Refs. \cite{stasto,sandapen,predazzi}). On the other hand, for vector mesons, the light-cone wavefunctions are not known
in a systematic way and they are thus obtained through models.  Here, we follows the
analytically simple DGKP approach \cite{dgkp:97}, which is found to describe in
good agreement vector meson production as pointed out in Ref.
\cite{sandapen}. In this particular approach, one assumes
that the dependencies on $\rr$ and $z$ of the wavefunction are
factorised, with a Gaussian dependence on $\rr$ (For a detailed discussion see Refs. \cite{sandapen,magno_victor_mesons}).

 Finally, the imaginary part of the forward amplitude can be obtained by
 putting the expressions for photon and vector meson (DGKP) wavefunctions into
 Eq. (\ref{sigmatot}). Moreover, summation over the quark/antiquark
 helicities  and an average over the   transverse polarization states
 of the photon should be taken into account. The transverse component (the longitudinal one does not contribute for photoproduction) is  then written as \cite{sandapen, dgkp:97}
\begin{eqnarray}
{\cal I}m \, {\cal A}_{T} (s,\,t=0) & = &   \int
d^{2}\rr \int_{0}^{1} dz \,\alpha_{{\mathrm{em}}}^{1/2} \,f_{V} \,
f_T(z)\,\exp \left[\frac{-\omega_T^{2}\,\rr^{2}}{2}\right] \nonumber \\
& \times & \left\{\frac{\omega_T^{2}\,\varepsilon \,r}{m_{V}}[z^{2} + (1-z)^{2}]
\,K_{1}(\varepsilon r) + \frac{m_{f}^{2}}{m_{V}}K_{0}(\varepsilon r)
\right \}\sigma_{dip}^{\mathrm{target}}(\tilde{x},\rr) \;,
\label{ampT}
\end{eqnarray}
with $\sigma_{dip}^{\mathrm{target}}$ being the dipole-proton
cross section in the nucleon case and the dipole-nucleus cross
section for scattering on nuclei. In the photoproduction case, $\varepsilon= m_f$, where $m_f$ is the quark mass of flavour $f$. The corresponding parameters for the vector mesons wavefunctions ($m_V$, $\omega_T$, $f_V$, etc) are presented in Table 1 of Ref. \cite{magno_victor_mesons}.  

In order to obtain the total cross section, we assume an exponential parameterization for the small $|t|$
behavior of the amplitude. After integration over $|t|$, the total
cross section for vector meson production by real/virtual photons in
the nucleon (proton) case reads as,
\begin{eqnarray}
\sigma\, (\gamma p \rightarrow Vp) = \frac{[{\cal I}m \, {\cal A}(s,\,t=0)]^2}{16\pi\,B_V}\,(1+\beta^{2}) \;
\label{totalcs}
\end{eqnarray}
where $\beta$ is the ratio of real to imaginary part of the
amplitude and $B_V$ labels the slope parameter. The values considered
for the slope parameter are taken from the  parameterization used
in Ref. \cite{mara}.  For the  $\rho$ case,  we
 have taken a different value in order to describe
 simultaneously  H1 and ZEUS photoproduction data.

In addition, Eq. (\ref{ampT}) represents only the
leading imaginary part of the positive-signature amplitude, and its
real part can be restored using dispersion relations ${\cal R}e \,{\cal
  A}=\tan (\pi \lambda/2)\,{\cal I}m {\cal
  A}$. Thus, for the $\beta$ parameter we have used the
simple ansatz,
\begin{eqnarray}
\beta = \tan \left(\frac{\pi\lambda_{\mathrm{eff}}}{2}
\right)\,,\hspace{0.7cm} \mbox{where}\;\;\;\lambda_{\mathrm{eff}}
= \frac{\partial \,\ln\,
  [{\cal I}m \, {\cal A}(s,\,t=0)]}{\partial \,\ln s}\,,
\end{eqnarray}
with $\lambda_{\mathrm{eff}}=\lambda_{\mathrm{eff}}(W_{\gamma
N},Q^2)$  the effective power of the imaginary amplitude, which
depends on both energy and photon virtuality. The correction
coming from  real part in  photoproduction, where only
transverse component contributes, is about  3\% for light mesons
and it reaches 13\% for $J/\Psi$ at high energies. It is worth mentioning that a different computation of the $\beta$ parameter, as in Ref. \cite{sandapen}, produces a larger effect even in the photoproduction case. An
additional correction is still required for heavy mesons, like
$J/\Psi$. Namely,  skewedness effects which takes into account the
off-forward features of the process (different transverse momenta
of the exchanged gluons in the $t$-channel),   are increasingly
important in this case.  Here, we follow  the studies in Ref.
\cite{Shuvaev:1999ce}, where the ratio of off-forward to forward
gluon  distributions reads as \cite{Shuvaev:1999ce},
\begin{eqnarray}
R_{g}\,(\lambda_{\mathrm{eff}})=\frac{2^{2\lambda_{\mathrm{eff}} + 3}}{\sqrt{\pi}}\,\frac{\Gamma\,\left(\lambda_{\mathrm{eff}}+ \frac{5}{2}\right)}{\Gamma \,\left(\lambda_{\mathrm{eff}}+4 \right)}\,,
\label{skew}
\end{eqnarray}
and we will multiply the total cross section by the factor $R_g^2$ for
the heavy meson case.

In the case of nuclear targets, $B_V$ is dominated by the nuclear
size, with $B\sim R_A^2$  and the non-forward differential cross section is
dominated by the nuclear form factor, which is the Fourier
transform of the nuclear density profile. Here we use the
analytical approximation of the Woods-Saxon distribution as a hard
sphere, with radius $R_A$, convoluted with a Yukawa potential with
range $a=0.7$ fm. Thus, the nuclear form factor reads
as \cite{klein_nis_prc},
\begin{equation}
F(q=\sqrt{|t|}) = \frac{4\pi\rho_0}{A\,q^3}\,
\left[\sin(qR_A)-qR_A\cos(qR_A)\right]
\,\left[\frac{1}{1+a^2q^2}\right]\,\,, \label{FFN}
\end{equation}
where $\rho_0 = 0.16$ fm$^{-3}$. Consequently, the photonuclear cross section is given by
\begin{eqnarray}
\sigma\,(\gamma A\rightarrow VA) =  \frac{[{\cal I}m \, {\cal
      A}_{\mathrm{nuc}}(s,\,t=0)]^2}{16\pi}\,\,(1+\beta^{2})\,\int_{t_{min}}^\infty
      dt\, |F(t)|^2 \,,
\label{fotonuclear}
\end{eqnarray}
with $t_{min}=(m_V^2/2\,\omega)^2$, where $\omega$ is the photon energy.

 Having introduced the main expressions
for  computing vector meson production in the color dipole
approach, in what follows we present a brief review of the saturation model and its
extension for the scattering on nuclei targets.
 In the present work, we follow the
quite successful saturation model \cite{golecwus}, which interpolates between the small and large dipole configurations,
providing color transparency behavior, $\sigma_{dip}\sim \rr^2$,
as $\rr \rightarrow 0$ and constant behavior, $\sigma_{dip}\sim
\sigma_0$, at large dipole separations. The parameters of the
model have been obtained from an adjustment to small $x$ HERA
data.   The parameterization for the dipole cross
section takes the eikonal-like form \cite{golecwus},
\begin{eqnarray}
\sigma_{dip}^{\mathrm{proton}} (\tilde{x}, \,\rr^2)  =  \sigma_0 \,
\left[\, 1- \exp \left(-\frac{\,Q_{\mathrm{sat}}^2(\tilde{x})\,\rr^2}{4} \right) \, \right]\,, \hspace{1cm} Q_{\mathrm{sat}}^2(\tilde{x})  =  \left( \frac{x_0}{\tilde{x}}
\right)^{\lambda} \,\,\mathrm{GeV}^2\,,
\label{gbwdip}
\end{eqnarray}
where the saturation scale $Q_{\mathrm{sat}}^2$ defines the onset of the
saturation phenomenon, which depends on energy. The parameters, obtained from a fit to the small-$x$ HERA data, are $\sigma_0=23.03 \,(29.12)$ mb, $\lambda= 0.288 \, (0.277)$ and
$x_0=3.04 \cdot 10^{-4} \, (0.41 \cdot 10^{-4})$ for a 3-flavor
(4-flavor) analysis. An
additional parameter is the effective light quark mass, $m_f=0.14$
GeV, which plays the role of a regulator for the photoproduction
($Q^2=0$) cross section.  The charm quark mass is
considered to be $m_c=1.5$ GeV. A smooth transition to the
photoproduction limit is obtained via the scaling variable \cite{golecwus}, $\tilde{x}= [(Q^2 + 4\,m_f^2)/(Q^2 + W_{\gamma N}^2)]$. 

The
saturation model is suitable in the region below $x=0.01$ and the
large $x$ limit needs still a consistent  treatment. Making use of
the dimensional-cutting rules, here we supplement
 the dipole cross section, Eq. (\ref{gbwdip}), with a threshold factor
$(1-x)^{n_{\mathrm{thres}}}$, taking $n_{\mathrm{thres}}=5$ for a
3-flavor analysis and $n_{\mathrm{thres}}=7$ for a 4-flavor one.
This procedure ensures consistent description of heavy quark
production at the fixed target data \cite{Mariotto_Machado}.

Let us discuss the extension of the saturation  model for the
photon-nucleus interactions. Here, we follow the simple procedure
proposed in Ref. \cite{armesto}, which consists of an extension to
nuclei of the
saturation model discussed above, using the Glauber-Gribov picture \cite{gribov}, without any new parameter. In this
approach, the nuclear version is obtained replacing the
dipole-nucleon cross section in Eq. (\ref{sigmatot}) by the
nuclear one,
\begin{eqnarray}
\sigma_{dip}^{\mathrm{nucleus}} (\tilde{x}, \,\rr^2;\, A)  = 2\,\int d^2b \,
\left\{\, 1- \exp \left[-\frac{1}{2}\,T_A(b)\,\sigma_{dip}^{\mathrm{proton}} (\tilde{x}, \,\rr^2)  \right] \, \right\}\,,
\label{sigmanuc}
\end{eqnarray}
where $b$ is the impact parameter of the center of the dipole
relative to the center of the nucleus and the integrand gives the
total dipole-nucleus cross section for a  fixed impact parameter.
The nuclear profile function is labeled by $T_A(b)$, which will
be obtained from a 3-parameter Fermi distribution for the nuclear
density  \cite{devries}. The above equation sums up all the
multiple elastic rescattering diagrams of the $q \overline{q}$
pair and is justified for large coherence length, where the
transverse separation $\rr$ of partons in the multiparton Fock state
of the photon becomes as good a conserved quantity as the angular
momentum, {\it i. e.} the size of the pair $\rr$ becomes eigenvalue
of the scattering matrix. It is important to emphasize that for
very small values of $x$, other diagrams beyond the multiple
Pomeron exchange considered here should contribute ({\it e.g.}
Pomeron loops) and a more general approach for the high density
(saturation) regime must be considered. However, we believe that 
this approach allows us to obtain lower limits of the high density 
effects. Therefore, the region of applicability of this  model should be at small values of  $x$, i.e. large coherence length, and for
not too high  values of virtualities, where the implementation of
the DGLAP evolution should be required. Consequently, the approach
is quite suitable for the analysis of exclusive vector meson
photoproduction in the   kinematical range of the planned
lepton-nucleus colliders (eRHIC and HERA-A) as well as UPC's at RHIC and LHC colliders. It is noticeable that the energy dependence of the cross sections is strongly connected with the saturation scale $Q_{s\,A}(W_{\gamma\,N})$. Namely, the saturation effects are larger  whether the momentum scale is of order or larger than the correspondent size of the vector meson and the energy growth of the cross section is then slowed down.

\section{Results and discussions}
\label{discussion}

In this section we  present the numerical calculation of the
rapidity  distribution and total cross section for the photoproduction of vector mesons in ultraperipheral heavy ion and proton - proton collisions.
Our main goal is to obtain  estimates from the QCD saturation model for vector meson photoproduction  in the  kinematical range of the colliders RHIC, LHC and Tevatron. Furthermore, a discussion concerning the comparison of the present results with the currently available models \cite{klein_nis_prc,strikman} based on vector meson dominance (VDM) and parton-hadron duality is also presented.  

\begin{figure}[t]
\begin{tabular}{cc}
\psfig{file=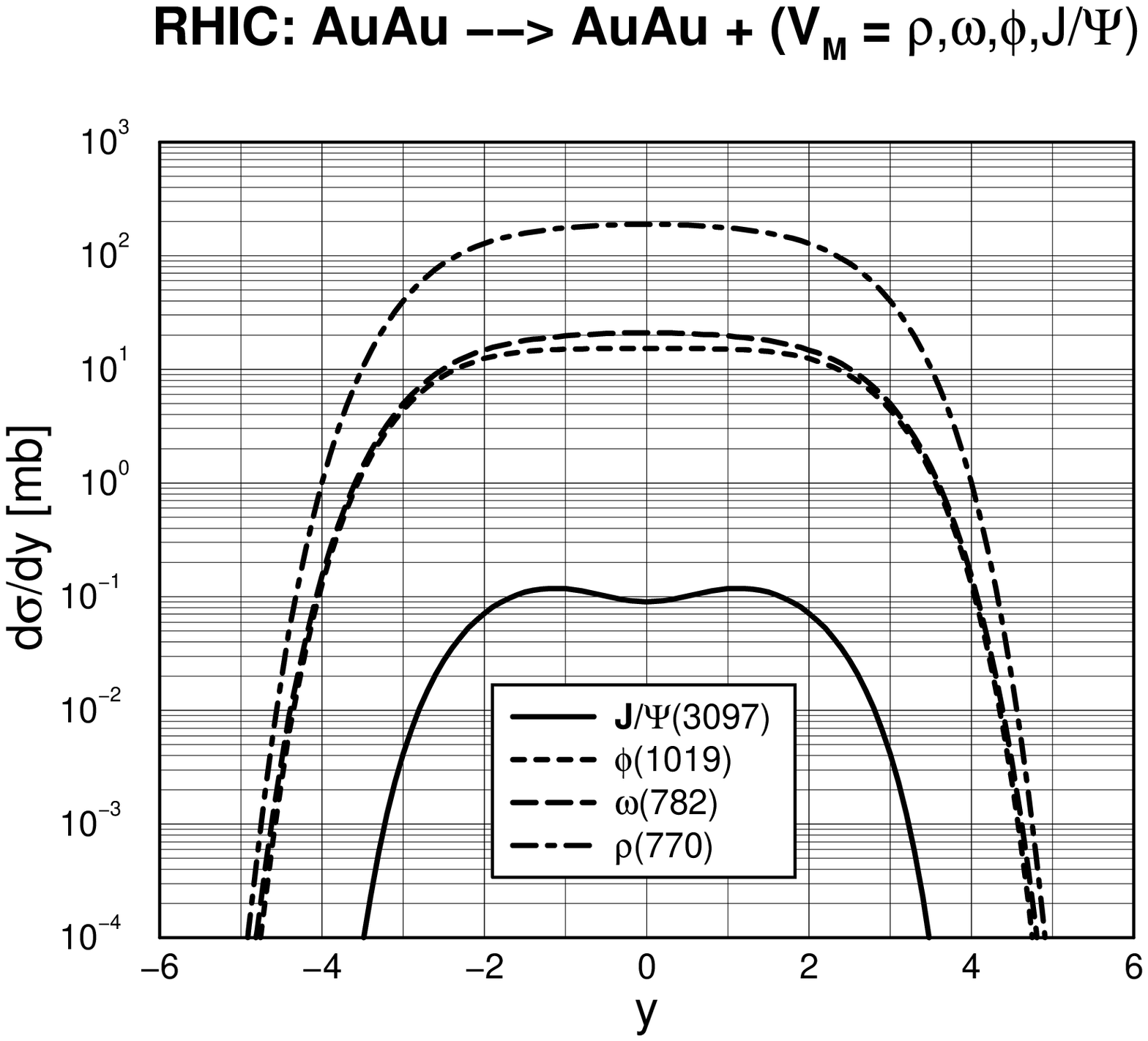,width=80mm} &
\psfig{file=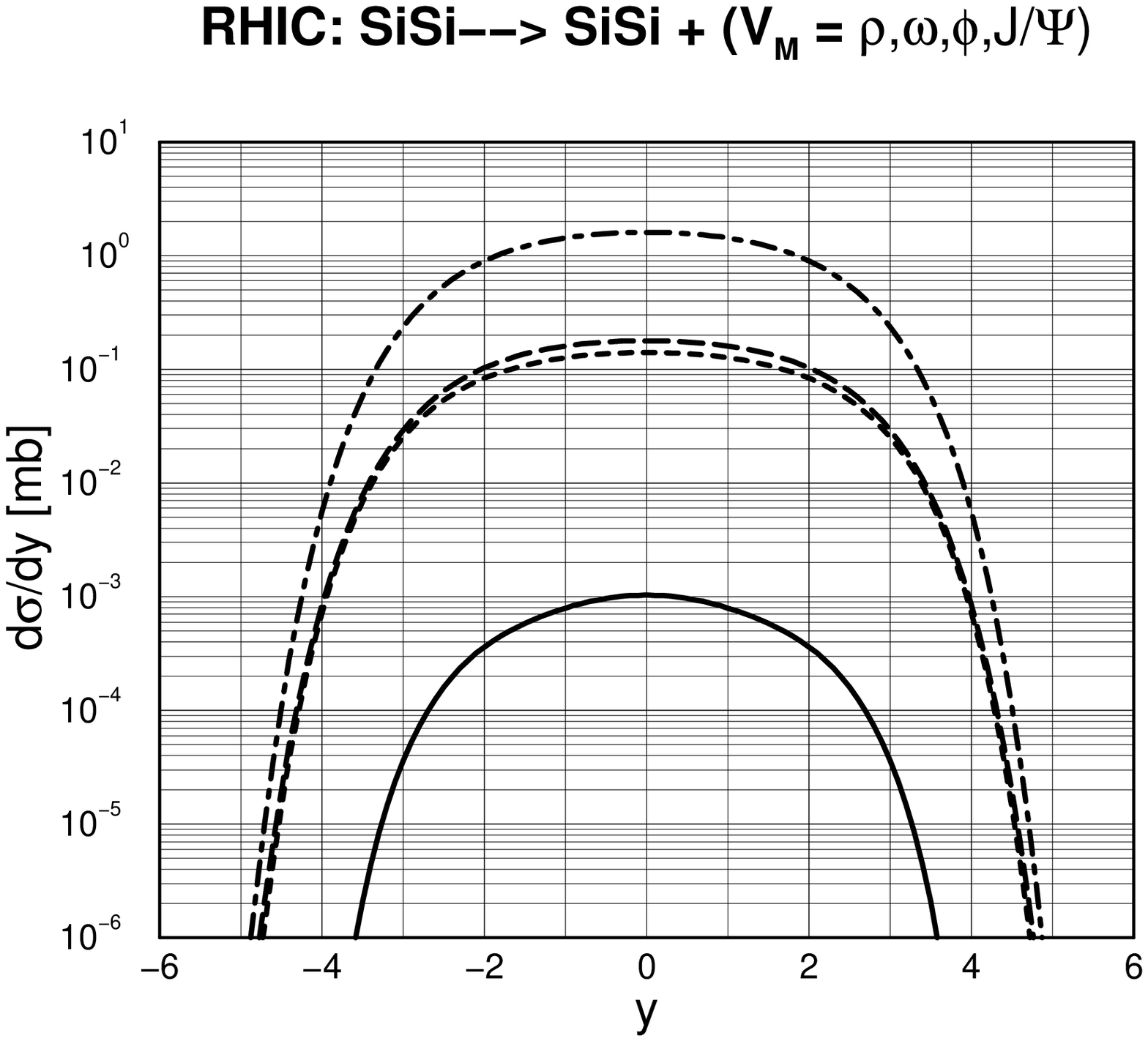,width=76mm}
\end{tabular}
 \caption{\it The rapidity distribution for nuclear vector meson  photoproduction on UPC's in $A A$  reactions at RHIC  energy ($\sqrt{S_{NN}}=0.2\,\mathrm{TeV}$).}
\label{fig1}
\end{figure}

\begin{figure}[t]
\begin{tabular}{cc}
\psfig{file=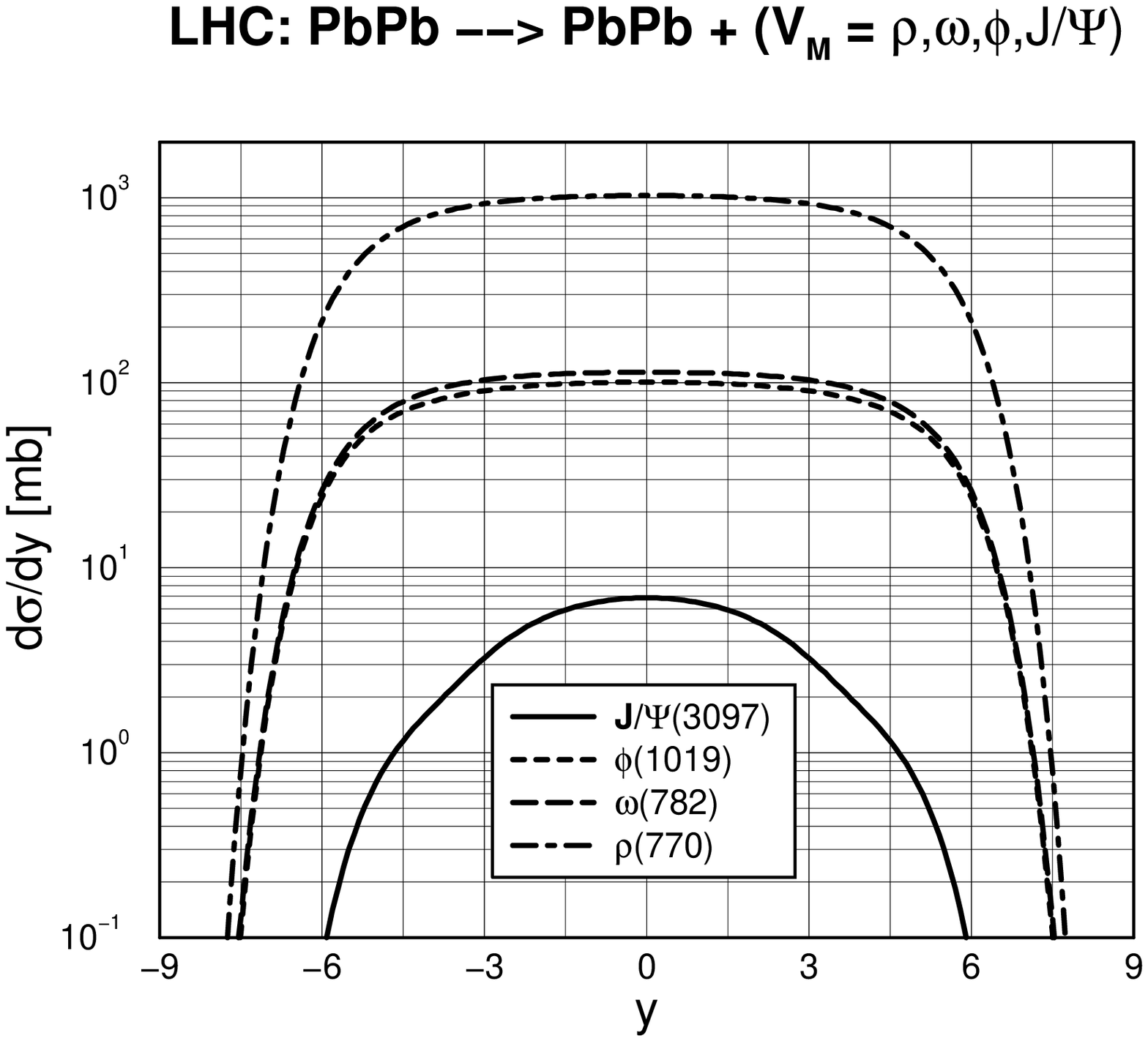,width=80mm} &
\psfig{file=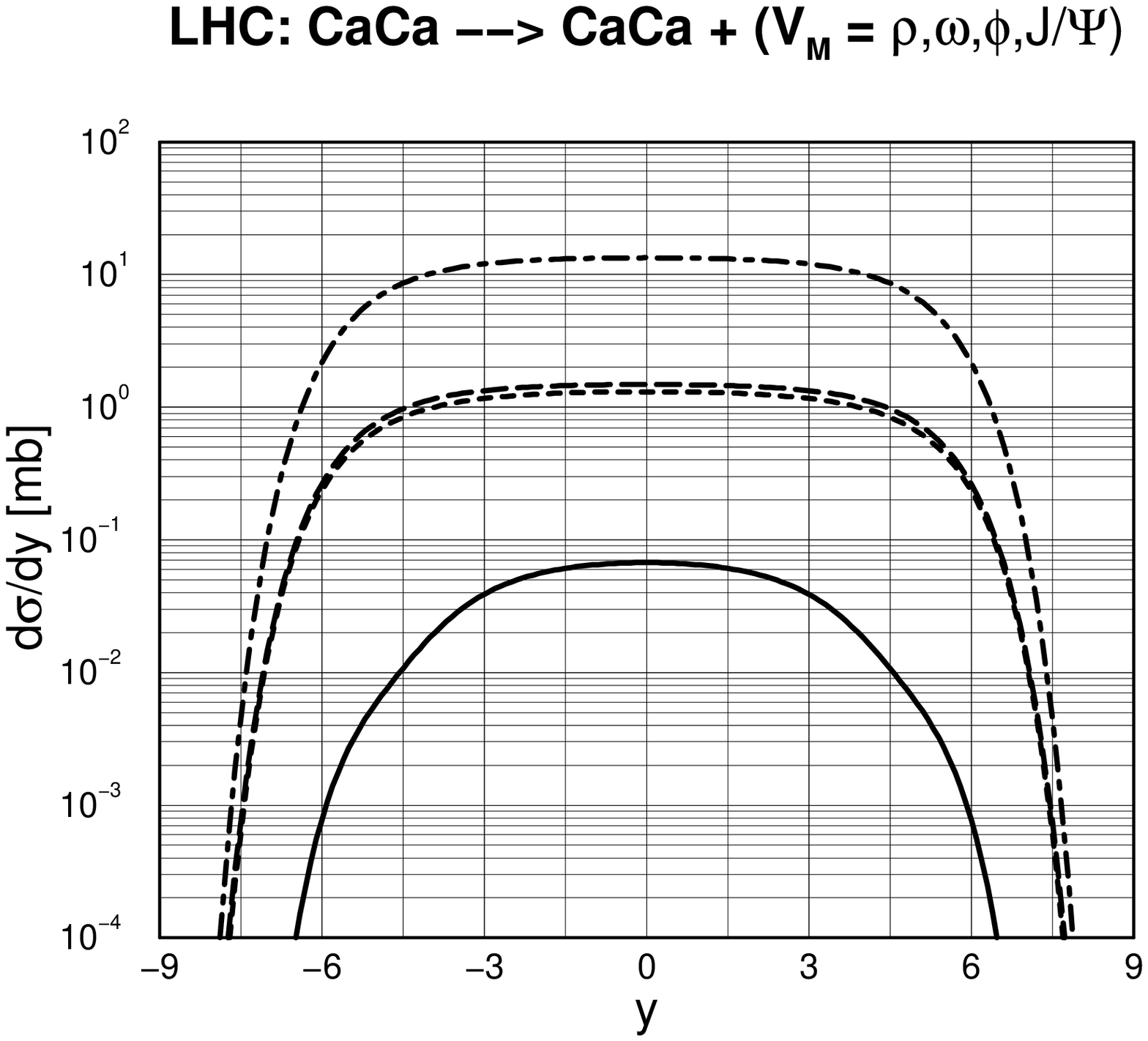,width=80mm}
\end{tabular}
 \caption{\it The rapidity distribution for vector meson  photoproduction on $A A$  reactions at LHC  energy $\sqrt{S_{NN}}=5.5\,\mathrm{TeV}$.}
\label{fig2}
\end{figure}

Initially, let us consider the nuclear photoproduction of vector mesons in ultraperipheral heavy ion collisions at RHIC ($\sqrt{s} = 200$ GeV) and LHC ($\sqrt{s} = 5500$ GeV) energies for different nuclei. In Figs. \ref{fig1} and \ref{fig2} one presents the predictions for meson rapidity distributions and in  Table \ref{tab1} one shows the corresponding integrated cross section. For LHC the considered nuclei are lead (Pb) and calcium (Ca), whereas for RHIC one takes gold (Au) and silicon (Si). The distributions for light and heavy vector mesons are placed in the same plot for sake of comparison among their orders of magnitude.  The final state rapidity is determined by the simple relation $y = \frac{1}{2}\ln\frac{\omega}{\sqrt{|t_{min}|}} = \ln \frac{ 2\,\omega}{m_{V}}$, that is the ratio between the photon energy $\omega$ and the longitudinal energy transfer in the laboratory frame. Consequently, the rapidity distribution reads as,
\begin{eqnarray}
\frac{d \sigma\,(AA\rightarrow AAV)}{d y} = \omega \, \frac{d \sigma\,(AA\rightarrow AAV)}{d \omega} \,\,,
\label{equpc}
\end{eqnarray}
where $\frac{d \sigma}{d \omega}$ is calculated using Eq. (\ref{sigAA}). Interchanging the photon emitter and target corresponds to a reflection around $y = 0$, with the total cross section being  the sum of these two possibilities. 

Let us discuss the numerical results. The rapidity distribution is characterized by a plateau at mid-rapidity at both RHIC and LHC, even for the light mesons as  the $\rho$ production. Such plateau is more pronounced at LHC. This is in contrast with the theoretical predictions of Refs. \cite{klein_nis_prc,strikman}, where due to the matching between the photon spectrum and the photonuclear cross section at lower (RHIC) energies a clear double-peak structure appears. For LHC, that feature disappears since lower photon energies contribute in a smaller amount to the total distribution. This  behavior is not present in our results mostly due to the inclusion of the threshold factor in the dipole-nucleus cross section, which enforces the correct behavior at energies near threshold region. This factor also prevents the sharp cut-off at the fragmentation (forward rapidities) region, which has appeared in recent calculation considering the $J/\Psi$ production \cite{klein_nis_prl}.  

The integrated cross sections can be contrasted with the theoretical estimations using GVDM plus Glauber-Gribov approach of Refs.\cite{strikman} as well as the estimation of Ref. \cite{klein_nis_prc}, which considers VDM plus a classical mechanical calculation for nuclear scattering  and uses as input for the $\gamma \, p \rightarrow V p$ reaction an extrapolation of the experimental DESY-HERA fits for meson photoproduction. Initially lets consider the latter approach (see Tab. III in Ref. \cite{klein_nis_prc}). At RHIC energy and Si nucleus, our results are about 20 \% lower for $\rho$ and $\omega$, whereas gives a larger $\phi$ cross section and almost the same $J/\Psi$ cross section. However, the situation changes for gold nucleus, where the present results are about 50 \% larger than the estimates in Ref. \cite{klein_nis_prc}. At LHC energy and for Ca nucleus, our results gives higher cross section by a factor of order 10 \%, whereas for lead nucleus the factor reaches a factor 2 for light mesons and almost similar for the $J/\Psi$ meson. Basically,  the values are quite similar for light nuclei. For heaviest nuclei, the results overestimate those ones in Ref. \cite{klein_nis_prc} when one considers light mesons and become similar for the $J/\Psi$ case. These features can be understood through the theoretical procedure when considering the nuclear scattering. 

\begin{table}[t]
\begin{center}
\begin{tabular} {||c|c|c|c|c|c||}
\hline
\hline
& {\bf HEAVY ION}   & $J/\Psi\,(3097)$ & $\phi\,(1019)$ & $\omega\,(782)$ & $\rho\,(770)$  \\
\hline
\hline
 {\bf RHIC} & SiSi &  3.42 $\mu$b & 612 $\mu$b & 764 $\mu$b &  6.74 mb \\
\hline
 & AuAu &  476 $\mu$b &  79 mb & 100 mb & 876 mb \\
\hline
\hline
 {\bf LHC} & CaCa &  436 $\mu$b & 12 mb & 14 mb &  128 mb \\
\hline
&  PbPb &  41.5 mb &  998 mb & 1131 mb & 10069 mb \\
\hline
\hline
\end{tabular}
\end{center}
\caption{\it The integrated cross section for nuclear vector mesons photoproduction at UPC's at RHIC and LHC energies.}
\label{tab1}
\end{table}

Despite the QCD color dipole model to be a different approach than that used in Ref. \cite{klein_nis_prc}, we can qualitatively to understand the discrepancy on the results by looking at the $\gamma \,A\rightarrow V A$ cross section in both models, which  is the  input for UPC's calculations in Eq. (\ref{sigAA}). For the dipole approach, this is given by Eq. (\ref{sigmatot}) replacing $p$ by the nucleus $A$. Qualitatively, one can roughly approximate the wavefunction as peaked at the dipole sizes around $\rr\sim 1/m_V=1/(2\,m_f)$ in the photoproduction case. Namely, $\Psi^\gamma\, \Psi^{V*}\propto \delta\,(r-1/m_V)\,\delta (z-1/2)$ and its normalization can be obtained from the electronic decay width constraint. This is supported by recent phenomenological studies on meson production within the dipole models \cite{sandapen,magno_victor_mesons,Nikolaev}.  Therefore, under this assumption the integration on longitudinal fraction $z$ and dipole size $r$ in Eq. (\ref{sigmatot}) can be carried out.  This procedure gives the following, up to a normalization factor,
\begin{eqnarray}
\frac{d \sigma_{\gamma\,A\rightarrow VA}^{\mathrm{dipole}}}{dt}|_{t=0}\propto \frac{\left[\sigma_{dip}^{\mathrm{nucleus}}\, (s, \,\rr^2=1/m_V^2;\, A) \right]^2}{16\pi}\,,
\end{eqnarray}
for the nuclear photoproduction of vector mesons.  The equivalent expression in the VDM approach is given by (See Eq. (13) in Ref. \cite{klein_nis_prc}), 
\begin{eqnarray}
\frac{d \sigma_{\gamma\,A\rightarrow VA}^{\mathrm{VDM}}}{dt}|_{t=0}\propto \frac{\left[\sigma_{tot}\, (VA) \right]^2}{16\pi}\,,
\end{eqnarray}
where the meson-nucleus total cross section has been labeled as $\sigma_{tot}\, (VA)$. In Ref. \cite{klein_nis_prc} 
the following expression has been assumed for this cross section (See Eq. (12) in Ref. \cite{klein_nis_prc})
\begin{eqnarray}
\sigma_{tot}\,(VA)  = \int d^2b 
\left\{\, 1- \exp \left[-\sigma_{tot}\,(Vp) \,T_A(b) \right] \, \right\}\,,
\label{CMexpr}
\end{eqnarray}
which drives the behavior on energy and atomic number of the photonuclear cross section. 
As discussed in detail in Ref. \cite{strikman} the above expression implies that in the black disk limit the total cross section becomes equal $\pi R_A^2$, which is the prediction of the classical mechanics. In contrast, a quantum mechanical approach implies that in that limit $\sigma_{tot} = 2 \pi R_A^2$. 
In Ref. \cite{strikman} the coherent $\rho$ production in UPC was studied using the generalized vector dominance model and the quantum mechanical Gribov-Glauber approach, which implies that   $\sigma_{tot}\, (VA) $ is given in the large coherence length limit by  
\begin{eqnarray}
\sigma_{tot}\,(VA)  = 2 \int d^2b 
\left\{\, 1- \exp \left[-\frac{1}{2}\sigma_{tot}\,(Vp) \,T_A(b) \right] \, \right\}\,.
\label{QMexpr}
\end{eqnarray}
On the other hand, in the QCD color dipole picture we have
\begin{eqnarray}
\sigma_{dip}^{\mathrm{nucleus}} \left(x,\, r\simeq \frac{1}{m_V}\right)  = 2\,\int d^2b \,
\left\{\, 1- \exp \left[-\frac{1}{2}\,\sigma_{dip}^{\mathrm{proton}} \left(x,\, r \simeq \frac{1}{m_V}\right) \,T_A(b) \right] \, \right\}\,,
\label{GBexpr}
\end{eqnarray}
where the dipole-proton cross section is given by Eq. (\ref{gbwdip}) for the saturation model. In particular, $\sigma_{dip}^{\mathrm{proton}} =\sigma_0\simeq 23-26$ mb in the case $Q_{\mathrm{sat}}^2(x)\,\gsim\, m_V^2$, while in the case where $Q_{\mathrm{sat}}^2(x)\ll m_V^2$ it presents the color transparency behavior $\sigma_{dip}\simeq Q_{\mathrm{sat}}^2/4m_V^2$. In what follows, we present a theoretical and numerical comparison among the approaches. 

The dipole-cross section is somewhat similar to the meson-hadron cross section, since it gives the probability of scattering of a color dipole (a $q\bar{q}$ pair) off a hadron. Therefore, we can consider them at a level of similarity. Hence, the main difference between Eqs. (\ref{CMexpr}) and (\ref{GBexpr}) comes from the rescattering procedure. In the dipole approach, one has the standard Glauber-Gribov  formalism, whereas in the VDM approach used in Ref. \cite{klein_nis_prc} a classical mechanical version has been used. In the color transparency regime, characteristic at intermediate energies (or light nuclei)  and/or heavy mesons, both the color dipole  and the meson-hadron cross section are small. This implies the first terms in the rescatterings are the leading ones and Eqs. (\ref{CMexpr}) and (\ref{GBexpr}) give quite similar results. On the other hand, at very high energies (or heavy nuclei)  and/or light mesons the black disk limit is reached. Thus, one obtains for the dipole approach $\sigma_{dip}^{\mathrm{nuc}}= 2\pi R_A^2$ and for the vector meson dominance model  $\sigma_{tot}(Vp)=\pi R_A^2$. In this limit case the ratio $\sigma_{\gamma A}^{\mathrm{dipole}}/\sigma_{\gamma A}^{\mathrm{VDM}}$ reaches a factor four. Similar arguments have been claimed in Ref. \cite{strikman}.

Now we compare our results with those ones in Ref. \cite{strikman}, where the main focus is on the $\rho$ and $J/\Psi$ production. For $\rho$ the predictions are computed only for RHIC energy $\sqrt{s_{NN}}=130$ GeV and we will consider it later on. We can anticipate that their results are closer to ours since a Glauber-Gribov approach is used in describing the scattering on nuclei. For $J/\Psi$ the theoretical approach for the photonuclear production was the collinear QCD double logarithmic approximation, where the $\gamma A \rightarrow J/\Psi A$ cross section is directly proportional to the squared nuclear gluon density distribution \cite{brodsky}. There, it was considered an impulse approximation (no nuclear shadowing) and a leading twist shadowing version. The impulse approximation gives a larger cross section at central rapidity (about a factor 4  higher for Ca and a factor 6 for Pb), while at fragmentation region both approximations match each other at LHC energy for Ca and Pb nuclei. Our results are closer to their impulse approximation, which suggests nuclear shadowing could be weak for $J/\Psi$ production. This feature can be easily tested in the first experimental measurements of coherent $J/\Psi$ production on UPC's at LHC. Concerning the integrated cross section, they found 0.6 mb for Ca nucleus and 70 mb for Pb at LHC. Our results are 0.44 mb and 41.5 mb, respectively. Thus, our results are about 15 \% lower for Ca and also 40 \% lower for Pb. The difference between the predictions comes from mostly from the distinct QCD approaches considered  used and the  different photon flux in the UPC calculation.

\begin{figure}[t]
\begin{tabular}{cc}
\psfig{file=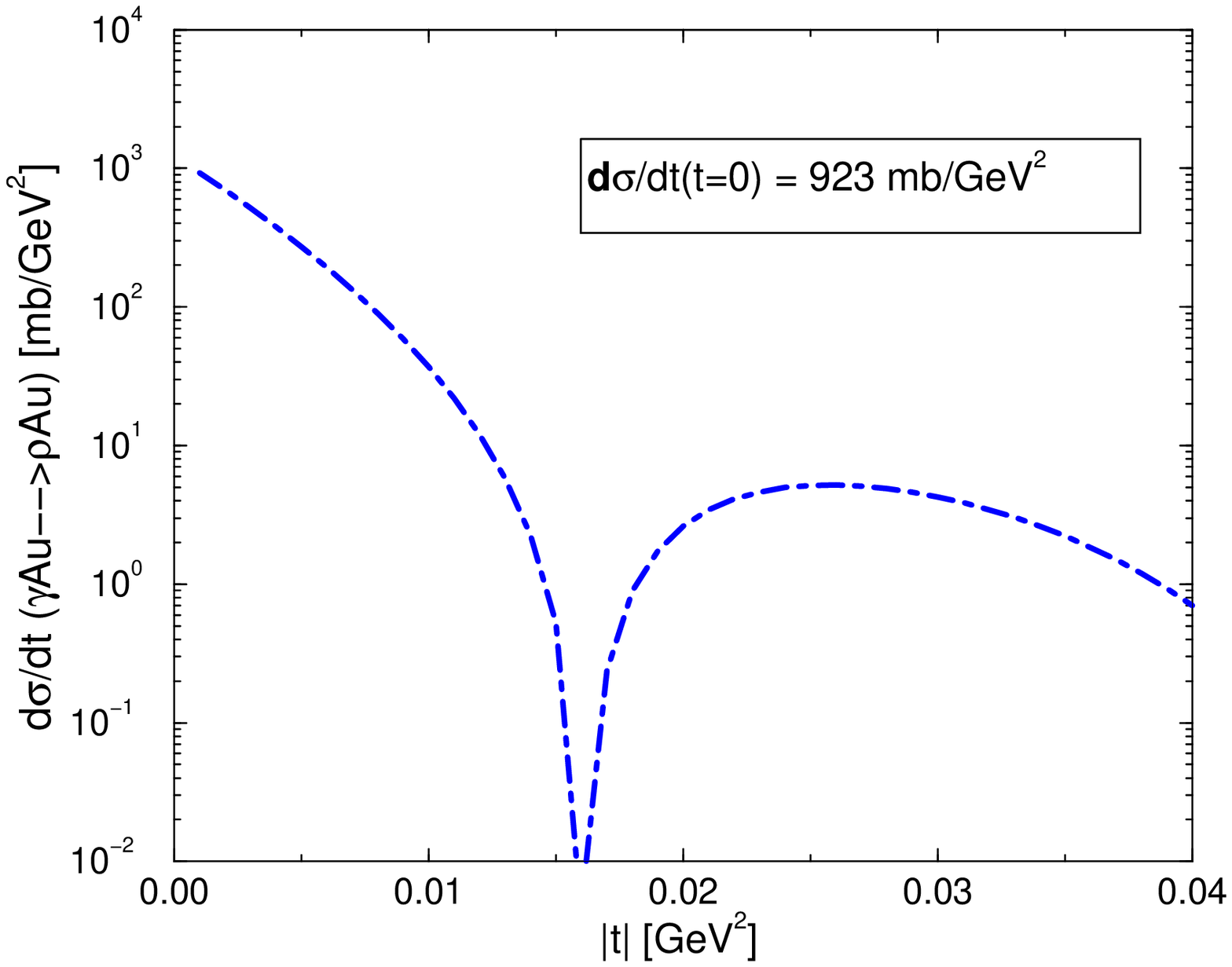,width=78mm} &
\psfig{file=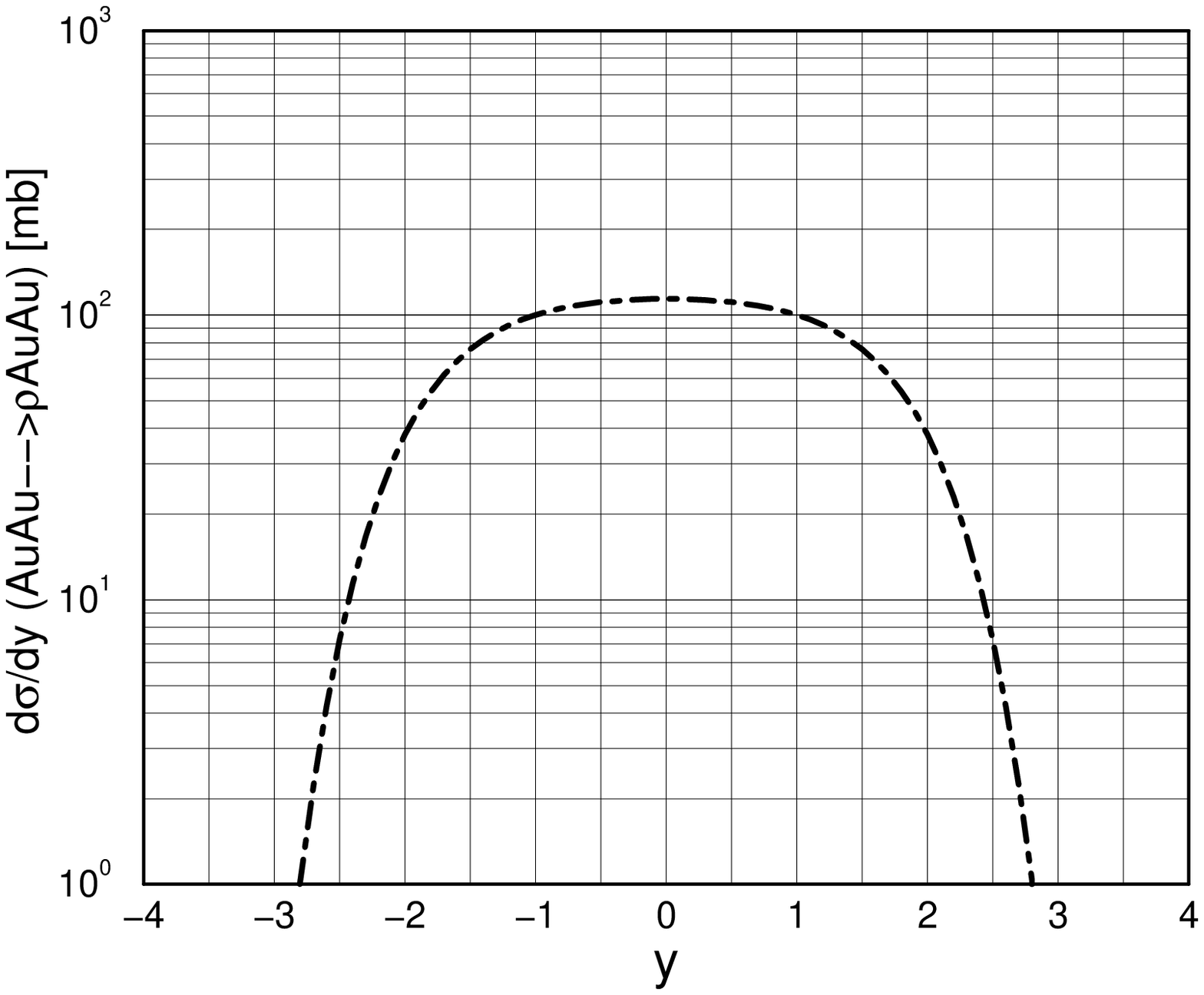,width=76mm}\\
(a) & (b)
\end{tabular}
 \caption{\it (a) Momentum transfer dependence of the $\rho$ meson photonuclear production at RHIC; (b) Rapidity distribution of coherent $\rho$ meson  production in gold-gold collisions at RHIC ($\sqrt{s_{NN}}=130$ GeV) for QCD dipole picture and saturation model.}
\label{fig3}
\end{figure}

Recently, the STAR Collaboration published  the first experimental measurement on the cross section of the coherent $\rho$ production in gold - gold ultraperipheral collisions at $\sqrt{s} = 130$ GeV \cite{star_data}, providing the first opportunity to test the distinct approaches describing nuclear vector meson photoproduction. In what follows we compare our results with these experimental data and confront them with other theoretical predictions currently available. In Fig. \ref{fig3}-(a) we present the momentum transfer behavior for the photonuclear $\rho$ production, which is the input for the UPC calculation. It is obtained by unfolding the integration over $|t|$ in Eq. (\ref{fotonuclear}). The dependence is proportional to the squared  of the nuclear form factor in Eq. (\ref{FFN}). The STAR Collaboration performed an exponential fit $d\sigma_{\gamma A\rightarrow \rho A}/dt\propto e^{-B_{\mathrm{nuc}}|t|}$ for this reaction, with an approximate gold radius of $R_{Au}=\sqrt{4B_{\mathrm{nuc}}}=7.5 \pm 2$ fm,  and obtained a forward cross section $ d\sigma_{\gamma A}(t=0)/dt=965\pm 140 \pm 230$ mb/GeV$^2$. Accordingly, we have found the value $d\sigma_{\gamma A}^{\mathrm{sat}}(t=0)/dt=923$ mb/GeV$^2$, which is consistent with the experimental results. The usual dip occurs at $|t|\simeq 0.015$ GeV, which  was also found in the calculation of Ref. \cite{strikman}. The rapidity distribution of the coherent $\rho$ production on UPC's is shown in Fig. \ref{fig3}-(b). The plateau at mid-rapidity remains, in contrast with the double-peaked structure appearing in the calculations of Ref. \cite{strikman}. However, the values at central rapidities are similar and of order $d\sigma (y\approx 0)/dy \simeq 100$ mb.

\begin{figure}[t]
\centerline{\psfig{file=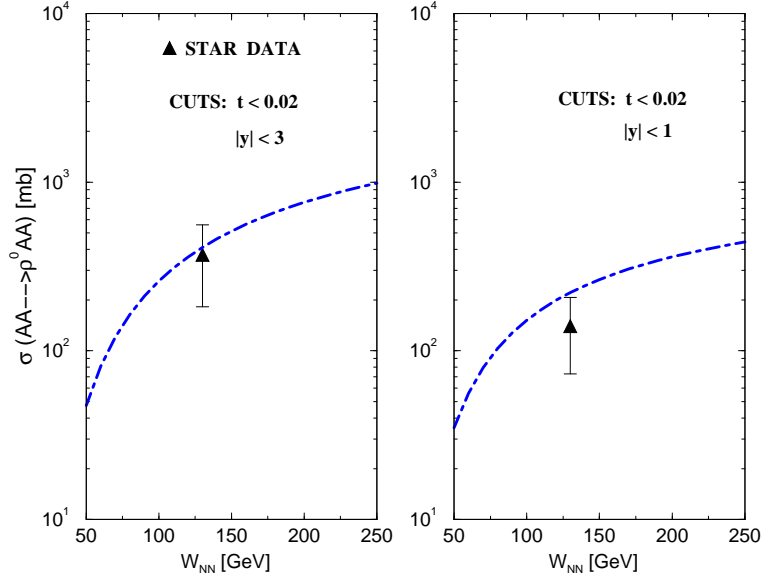,width=100mm}} 
 \caption{\it Energy dependence of coherent $\rho$ meson  production in gold-gold in UPC's at RHIC ($\sqrt{s_{NN}}=130$ GeV) in the QCD dipole picture and saturation model. Experimental data from STAR Collaboration \cite{star_data}.}
\label{fig4}
\end{figure}

The energy dependence of the cross section for the photoproduction of the meson $\rho$ in gold-gold ultraperipheral  collisions is presented in Fig \ref{fig4}. For comparison, the experimental data from STAR Collaboration at RHIC at $\sqrt{s_{NN}} = 130$ GeV \cite{star_data} is also shown. Our theoretical estimations in the curves take into account the experimental cuts. The cut on the momentum transfer $|t|<0.02$ GeV$^2$ slightly reduces the cross section by a few percents. In the range of rapidities $|y|\leq 3$ (left plot), at energy $\sqrt{s_{NN}}=130$ GeV, we have found $\sigma_{\mathrm{sat}}(-3\leq y \leq 3)= 410$ mb in good agreement with the STAR measurement $\sigma_{\mathrm{STAR}}(-3\leq y \leq 3)= 370\pm 170 \pm 80$ mb. For the cut  $|y|\leq 1$ (left plot), we have obtained $\sigma_{\mathrm{sat}}(-1\leq y \leq 1)= 221$ mb, whereas the STAR result is $\sigma_{\mathrm{STAR}}(-1\leq y \leq 1)= 140\pm 60 \pm 30$ mb. In this case, our result is about 35 \% higher than the central value of the STAR measurement. The values presented here are somewhat similar to the ones obtained in Ref \cite{strikman}, which uses the generalized vector dominance model (GVDM) and the Glauber-Gribov approach, including in addition the finite coherence length effects. As discussed before, our calculation on the photonuclear cross section $\gamma A \rightarrow VA$ considers large coherence length. This is not the case at low photon energies released at RHIC and the finite length effect could further suppress the cross section. We believe this suppression should not be strong, once they obtained $\sigma_{\mathrm{GVDM}}(-3\leq y \leq 3)= 490$ mb and $\sigma_{\mathrm{GVDM}}(-1\leq y \leq 1)= 170$ mb, respectively.

\begin{figure}[t]
\begin{tabular}{cc}
\psfig{file=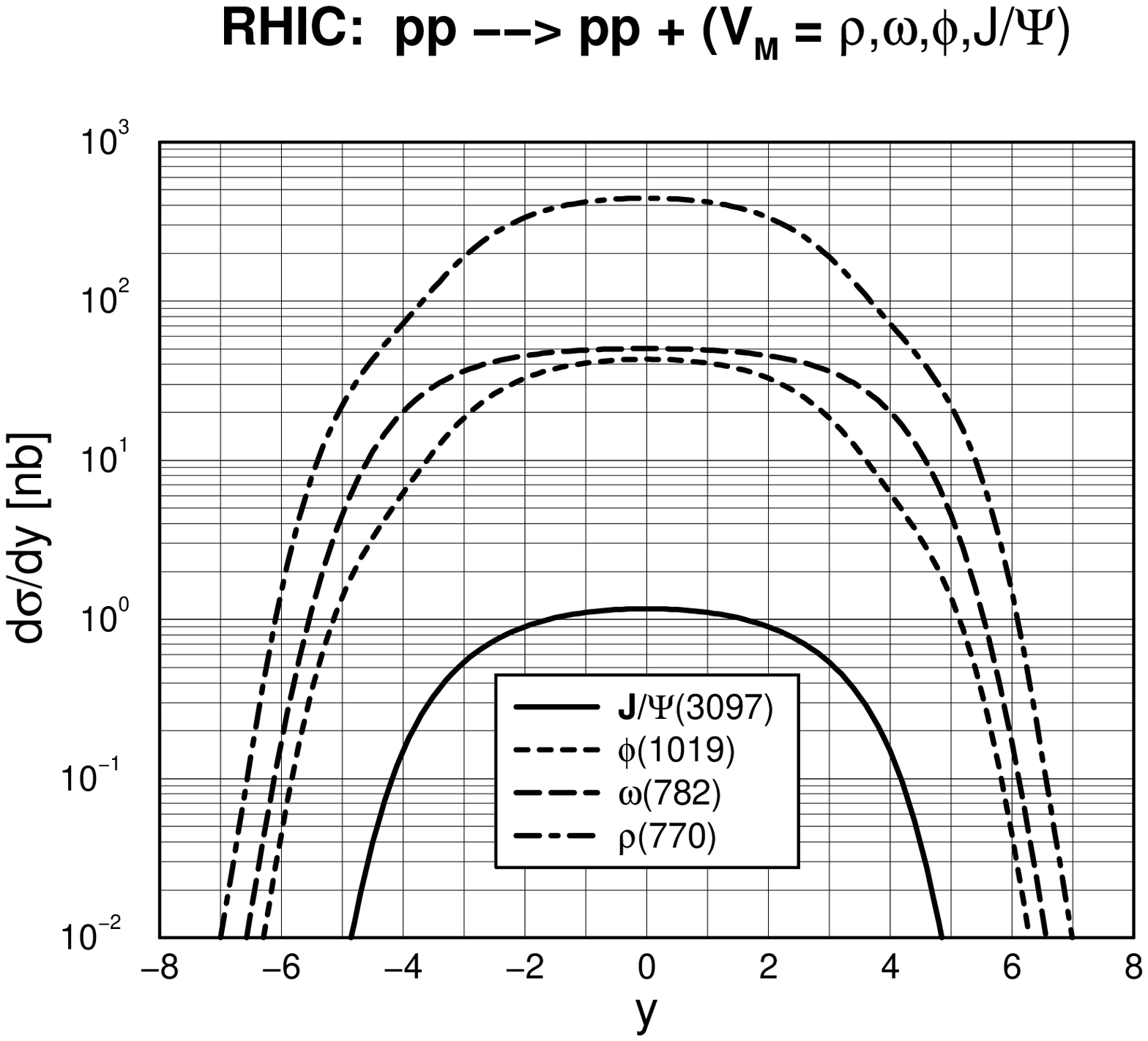,width=80mm} &
\psfig{file=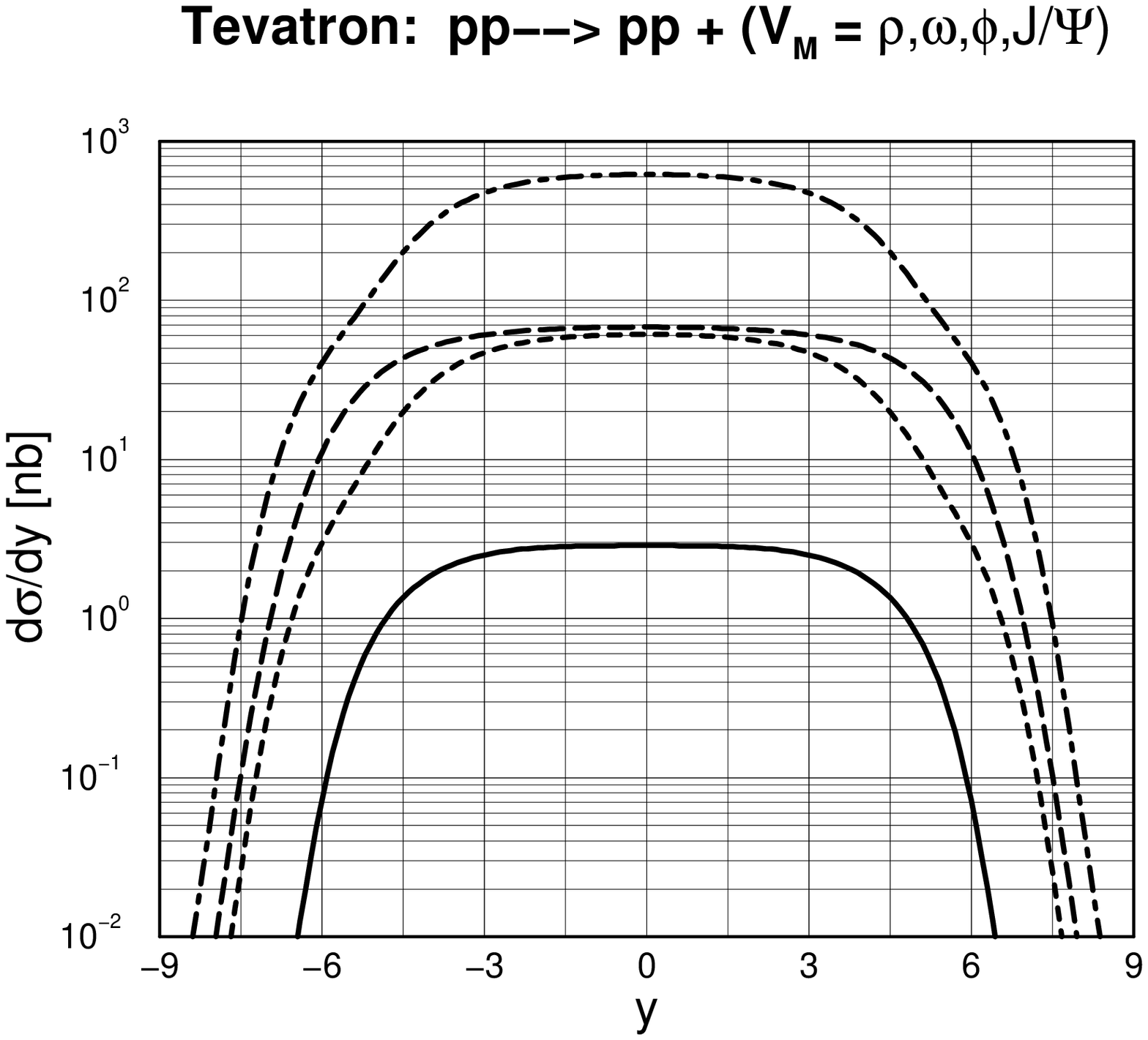,width=82mm}
\end{tabular}
 \caption{\it  The rapidity distribution for vector meson  photoproduction on $pp$ or $p\bar{p}$ reactions at RHIC ($\sqrt{S_{NN}}=0.5\,\mathrm{TeV}$) and Tevatron ($\sqrt{S_{NN}}=1.96\,\mathrm{TeV}$) energies. }
\label{fig5}
\end{figure}

\begin{figure}[t]
\centerline{\psfig{file=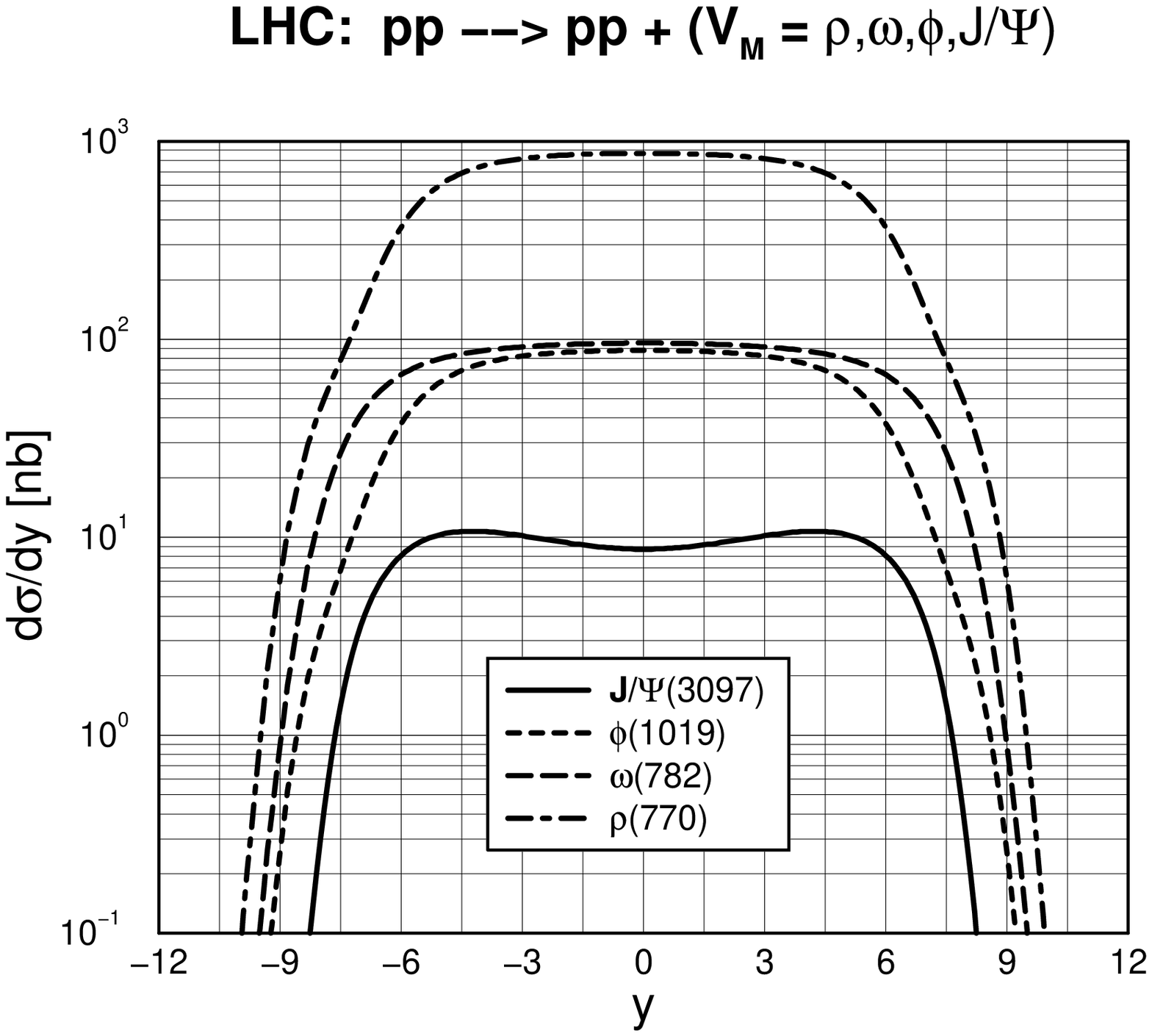,width=80mm}}
 \caption{\it  The rapidity distribution for vector meson  photoproduction on $pp$  reactions at LHC energy  $\sqrt{S_{NN}}=14 \,\mathrm{TeV}$.}
\label{fig6}
\end{figure}

Finally, lets now consider the photoproduction of vector mesons in proton - proton collisions  for RHIC ($\sqrt{s} = 500$ GeV), Tevatron ($\sqrt{s} = 1960$ GeV)  and LHC ($\sqrt{s} = 14000$ GeV) energies (For a similar analysis of the photoproduction of heavy quarks in $pp$ collisions see Ref. \cite{vicmag_prd}). In Figs. \ref{fig5} and \ref{fig6}, we present our predictions for the rapidity distributions and in the Table \ref{tab2} our results for the integrated cross section. The distribution on rapidity has similar features as for the nuclear case, presenting a plateau at mid-rapidity and without double-peaked structure. Concerning the integrated cross sections, for $J/\Psi$ production a comparison with the results obtained in the Ref. \cite{klein_vogt}  is possible. It is important to emphasize that in  that reference  a parameterization for the total $\gamma p \rightarrow J/\Psi p$ cross section is used as input in the calculations. In this case we have that our predictions  are similar for RHIC energies, being approximately $10 \%$ larger for Tevatron and LHC energies, which is expected since the saturation model describes reasonably well the HERA data. Very recently, predictions for the $\rho$ and $\phi$ photoproduction  in proton - proton collisions have been presented in Ref. \cite{nys_fluxo}. We have that our results for $\phi$ production are 20 $\%$ smaller for RHIC energies but similar for LHC energy. In contrast, we have that our results for $\rho$ production reach a factor 2 smaller than the predictions from \cite{nys_fluxo} for all energies. This difference  is due to the sizeable importance of the threshold factor. If this factor is not taken into account, the behavior of the cross section near threshold $\mu_{\mathrm{thres}}=(m_p+m_V)$ is overestimated. This gives additional contribution in an integrated cross section, mostly in the $\rho$ case where the threshold scale stays at low energies $\mu_{\mathrm{thres}}^ {\rho} \approx 1.6$ GeV. Namely, for light mesons, the differential cross section $d\sigma/dy$ receives a larger contribution from the low photon energies region which it is quite sensitive to the threshold corrections.

Finally, lets present a  short discuss on the experimental feasibility of the reactions considered here. Although the vector meson photoproduction at $AA$ or $pp$ collisions to be a small fraction in comparison to the total hadronic cross sections, the experimental separation of these reaction channels is possible. The rates are large even after taking into account the respective leptonic branching rations and/or acceptance estimates. The results presented in Tabs. \ref{tab1} and \ref{tab2} show that the coherent photoproduction of light mesons are very high at RHIC and LHC. For instance, the exclusive $\rho$ production coming from these reactions reaches 10\% of the total nucleus-nucleus cross section at RHIC, whereas corresponds to 50\% of the PbPb total cross section at LHC. As photoproduction is an exclusive reaction, $N+N \rightarrow N+N+V$ ($N=p,\,A$), the separation of the signal from hadronic background would be relatively clear. Namely, the characteristic features in photoproduction at UPC's are low $p_T$ meson spectra and a double rapidity gap pattern. Moreover, the detection (Roman pots) of the scattered nuclei (or protons) can be an additional useful feature. In hadroproduction, the spectra on transverse momentum of mesons are often peaked around meson mass, $p_T\approx m_V$. A experimental cut $p_T< 1$ GeV would eliminate most part of the hadronic background. Hence, the rapidity cut would enter as an auxiliary separation mechanism. This procedure is specially powerful, since there will be rapidity gaps on both sides of the produced meson. For numerical estimates on these cuts procedures, we quote Refs. \cite{klein_nis_prl,nys_fluxo}.

\begin{table}[t]
\begin{center}
\begin{tabular} {||c|c|c|c|c||}
\hline
\hline
 $\mathbf{V_M \,(m_V)}$/{\bf COLLIDER} & $J/\Psi\,(3097)$   & $\phi\,(1019)$  & $\omega\,(782)$ & $\rho\,(770)$\\
\hline
\hline
 {\bf RHIC} & 6.57 nb &  242.4 nb & 361.7 nb & 2.52 $\mu$b \\
\hline
\hline
{\bf Tevatron} & 24.64 nb & 476 nb & 646 nb & 4.84 $\mu$b \\
\hline
\hline
 {\bf LHC} & 132 nb &  980 nb & 1.24 $\mu$b & 9.75 $\mu$b \\
\hline
\hline
\end{tabular}
\end{center}
\caption{\it The integrated cross section for the photoproduction of vector mesons in $p+p(\bar{p})$  collisions at RHIC, Tevatron and LHC.}
\label{tab2}
\end{table}

\section{Summary}
\label{conclusions}

In summary, we have calculated the rapidity distribution and integrated cross sections of  exclusive photonuclear production of vector mesons in ultraperipheral heavy ion collisions within the QCD color dipole picture, with 
 particular emphasis on the saturation model. This is motivated by the good agreement of this model in systematically describing the current data on vector meson photoproduction in scattering on protons and its reliable estimates for scattering on nuclei.  The cross sections for the $A+A \rightarrow A+A+V$ ($V = \rho, \omega, \phi, J/\Psi$) process were computed and theoretical estimates for scattering on both  light and heavy nuclei are given for RHIC and LHC energies. The rates are very high, mostly for light mesons and at LHC energies. In particular, we compare our prediction for the coherent $\rho$ meson production with  RHIC data at $\sqrt{s_{NN}}=130$ GeV. The corresponding results are in good agreement with the experimental results when considered the cuts on momentum transfer and on rapidity. We also have contrasted our results with the current models which consider vector dominance (VDM) or generalized vector meson dominance (GVDM) and Glauber-Grivov formalism for the nuclear scattering. We have pointed out the main sources of the discrepancies among those models and our estimates. In addition, the cross sections for the $p+p(\bar{p}) \rightarrow p+p(\bar{p})V$  process are computed and theoretical estimates are given for the hadronic colliders RHIC and LHC in their $pp$ mode and Tevatron. Finally, the experimental feasibility and signal separation on the reaction channels presented here are shortly discussed. Although the rates are lower than hadroproduction, the coherent photoproduction signal would be clearly separated by applying a transverse momentum cut $p_T<1$ and two rapidity gaps in the final state.

\section*{Acknowledgments}
 One of us (M.V.T.M.) thanks the support of the High  Energy Physics Phenomenology Group, GFPAE IF-UFRGS, Brazil. This work was partially financed by the Brazilian funding agencies CNPq and FAPERGS.

\end{document}